\documentclass[layout,reprint,revtex4-1,prl,aps,superscriptaddress,showpacs]{revtex4-1}%reprint produces close to actual layout, preprint: gives double-spaced to edit
\usepackage{amsmath}    % need for subequations
\usepackage{graphicx}   % need for figures
\usepackage{verbatim}   % useful for program listings
\usepackage{color}      % use if color is used in text
\usepackage{subfigure}  % use for side-by-side figures
\usepackage{hyperref}   % use for hypertext links, including those to external documents and URLs
\newcommand{\un}[1]{\ensuremath{\, \mathrm{#1}}}

\begin{document}

\title{Control of synchronization patterns in neural-like Boolean networks}
\author{David P. Rosin}
\email[Electronic address:]{rosin@phy.duke.edu}
\affiliation{Department of Physics, Duke University, Durham, North Carolina 27708, USA}
\affiliation{Institut f\"ur Theoretische Physik, Technische Universit\"at Berlin, Hardenbergstra{\ss}e 36, 10623 Berlin, Germany}
\author{Damien Rontani}
\affiliation{Department of Physics, Duke University, Durham, North Carolina 27708, USA}
\author{Daniel J. Gauthier}
\affiliation{Department of Physics, Duke University, Durham, North Carolina 27708, USA}
\author{Eckehard Sch\"oll}
\affiliation{Institut f\"ur Theoretische Physik, Technische Universit\"at Berlin, Hardenbergstra{\ss}e 36, 10623 Berlin, Germany}

\date{\today}
%Nonlinear dynamical systems, neural networks, Control theory and feedback in neuroscience, 
\begin{abstract}
We study experimentally the synchronization patterns in time-delayed directed Boolean networks of excitable systems. We observe a transition in the network dynamics when the refractory time of the individual systems is adjusted. When the refractory time is on the same order-of-magnitude as the mean link time delays or the heterogeneities of the link time delays, cluster synchronization patterns change, or are suppressed entirely, respectively. We also show that these transitions occur when we only change the properties of a small number of driver nodes identified by their larger in-degree, hence the synchronization patterns can be controlled locally by these nodes. Our findings have implications for synchronization in biological neural networks.
\end{abstract}

\maketitle

\paragraph{Introduction}
%zero-lag cluster synchronization with delays is interesting, linked to cognition, patterns appear in the brain and are conjectured to cause brain functioning
One striking dynamical phenomenon arising in complex networks is the existence of zero-time-lag synchronized behavior in the presence of link time delays  \cite{ROE97,VIC08,FIS06,DAH12}. In neural networks, time delays result from propagation of neuronal pulses along the axons introducing several tens of milliseconds of latency, which is significantly larger than the duration of the action potential $(\lesssim 1\un{ms})$ \cite{RIN94}. Yet, even between distant parts of the brain, neural activity that is synchronized with zero time-lag has been observed \cite{ROE97,ROD99,FRI97a,SCH06i,Var01} and found to be associated with perception and neurological diseases \cite{SCH11e,UHL06}. These dynamics of repeating spiking patterns are responsible for a wide range of rhythmic motor behaviors \cite{MAR01,STE99,SEL10}, which are dependent on the network refractory time \cite{HAR12}.

%our structures appear in the brain, the GCD theory
Studies on neurological networks such as in C.~elegans found recurring topological structures of nodes assembled in loops with directed connections \cite{MIL02}. This type of network displays stable synchronization patterns with properties that can be inferred from the network topology alone. Specifically, cluster synchronization originates from the distribution of one initial stimulus (a pulse) in the network via the time-delayed links. At nodes that have multiple inputs, signals are combined and the propagation times from their output to their input via the different loops are given by the loop sizes and, in addition, are affected by heterogeneity in the link time delays. If the heterogeneity is negligible, the dynamics relaxes to a stable synchronization pattern of zero-lag synchronized clusters, where the number of clusters---groups of neurons that are individually zero-lag synchronized---is given by the greatest common divisor (GCD) of the number of nodes in each network loop \cite{KAN11a,KAN11,NIX12,VAR12,VAR12a}.

%limitations of the GCD theory and MSF
This prediction, however, relies on a separation of time scales, such that link time delays are larger than the characteristic time scale of the node dynamics. In addition, local variations of the coupling strength and noise can affect the synchronization patterns of the network, so that the 
non-local criterion of the GCD can fail to describe the synchronization state \cite{KOP12}.
In general, stability of group and cluster synchronization in multipartite networks, i.e., networks 
where each cluster is coupled unidirectionally to only one other cluster, can be predicted using the master 
stability approach \cite{DAH12}.
In the special case that a network of identical nodes consists of directed loops only, the GCD theory can 
be applied to calculate the largest possible number of clusters.

%overview
In this Letter, we study synchronization patterns in an experimental time-delayed network of excitable nodes for different regimes of the refractory times. First, we find a loss of synchronization when the refractory times are comparable to the heterogeneity of the link time delays. Second, we find a modification in the cluster synchronization patterns when the refractory times are larger than the link time delays. These effects can be controlled by adjusting the refractory times of only a few important nodes in the network that have the largest in-degree---a quantity that describes the number of input connections. The observed loss and dynamical modification of neural synchronization patterns through a variation of the refractory time might also appear in the brain, where it could be applied to control rhythmic motor behaviors and neurological diseases by adjusting the refractory time with chemicals in the blood \cite{KEN79}.

\paragraph{Setup}
We build Boolean networks with electronic logic gates using a field-programmable gate array (FPGA) because it offers a large number (${\sim}100{,}000$) of logic elements with reassignable logic functions and flexible interconnections. Our circuits operate autonomously so that their dynamics are governed by the logic gate propagation delays and connection time delays, and not by a global clock. The continuous temporal evolution can be described mathematically by both Boolean delay equations and ordinary differential equations \cite{GHI85,MES97,GLA98,EDW00}. Theoretically and experimentally, these autonomous Boolean networks have been found to display a large variety of dynamical behaviors, such as chaos, periodic oscillations, and excitability \cite{ZHA09a,CAV10,ROS12}.
 
\begin{figure}[tb]
\centering
\includegraphics{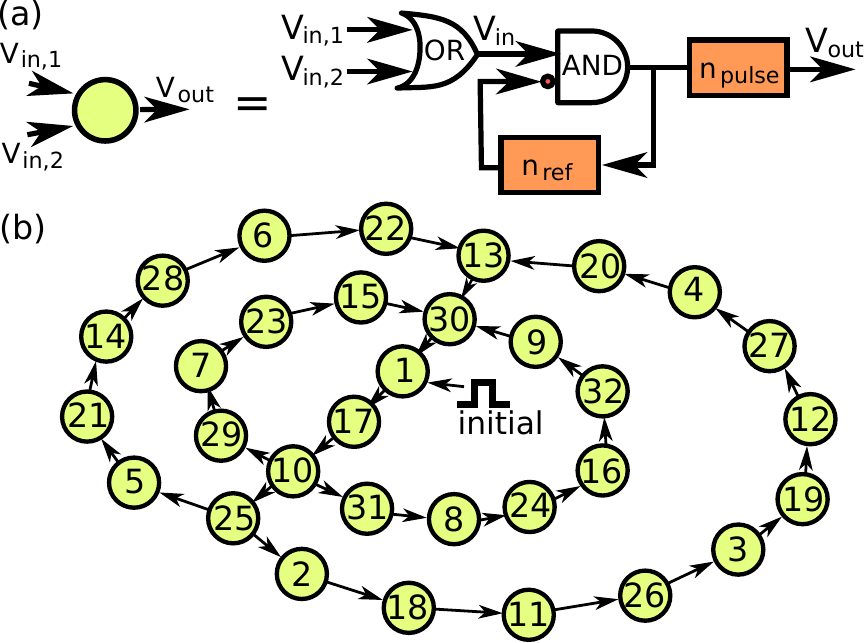}%[width=1\linewidth]
\caption{(Color online) 
(a)~Scheme of an excitable node and its implementation with autonomous logic gates using two pulse generators (boxes). An OR gate combines multiple inputs. A total number of $2(n_\mathrm{ref}+n_\mathrm{pulse})+3$ logic elements are required to implement an excitable node with one input.
(b)~The network topology, where directed time-delayed connections are indicated by arrows, nodes are indicated by circles and labeled to identify them in the following figures, and the node that receives the initial pulse is also indicated.
}
\label{fig:setup}
\end{figure}

Here, we realize excitable nodes with a circuit design of autonomous logic gates that is introduced in Ref.~\cite{ROS12} and shown schematically in Fig.~\ref{fig:setup}a. Similar to biological neurons \cite{KEE09}, the electronic circuit responds to above-threshold input signals with large output signals (pulses) and leaves its stable fixed point for a finite time \cite{IZH07}. It is governed by two characteristic time scales given by its output pulse width $T_\mathrm{pulse}$ and a quiescent time period after generating a pulse, which is the refractory time $T_\mathrm{ref}$ (mnemonic refractory). 

The two time scales $T_\mathrm{pulse}=n_\mathrm{pulse}\tau_\mathrm{buf}$ and $T_\mathrm{ref}=n_\mathrm{ref}\tau_\mathrm{buf}$ can be varied by changing the numbers $n_\mathrm{pulse}$ and $n_\mathrm{ref}$ of autonomous delaying elements (buffers, implemented with two consecutive inverter gates), where $\tau_\mathrm{buf}=(560\pm 20)\un{ps}$ is the propagation time through one delaying element. The quantity $\tau_\mathrm{buf}$ fluctuates in time by $\pm 1\%$ due to timing jitter. In addition, $\tau_\mathrm{buf}$ has slightly different values for different logic elements due to process variation in the fabrication of the integrated circuits, quantified by a heterogeneity of $\pm 3.5\%$. Here, we implement networks of several excitable nodes using many logic gates on the chip. We state average values for the properties of the nodes, which are therefore affected by the latter error estimation. 

We build networks by connecting excitable nodes with directed time-delayed links, where link time delays $\tau=n_\tau \tau_\mathrm{buf}$ are realized with $n_\tau$ cascaded delaying elements. When nodes have multiple input connections, these signals are combined with electronic equivalents of neurological synapses \cite{IND11}. We implement electronic synapses with OR gates so that any of the inputs can excite the node. In this fashion, we can create large networks and study the dynamics.

\paragraph{Synchronization patterns}
We study cluster synchronization in a network with a topology shown in Fig.~\ref{fig:setup}b with $N=32$ excitable nodes assembled in four directed loops of 8, 10, 12, and 16 elements. First, we consider the case of a separation of time scales with node refractory times of $T_\mathrm{ref}=(5.6\pm0.2)\un{ns}$. The $i$th loop has a propagation time $T_i$ associated with it, given by 
\begin{equation}\label{eq:loop_prop_time}
T_i=L_i(\tau+\delta)+\Delta_i, 
\end{equation}
where $L_i$ is the number of nodes in the loop, $\tau=(16.7\pm0.6)\un{ns}$ is the delay of a single link, $\delta$ is the processing delay of one node, and $\Delta_i$ is the heterogeneity in loop $i$. We measure the maximum heterogeneity in our network to be $\Delta=\max_{i,j}(\left|\Delta_i-\Delta_j\right|)=(2.8\pm0.1)\un{ns}$.

\begin{figure}[tb]
\centering
\includegraphics{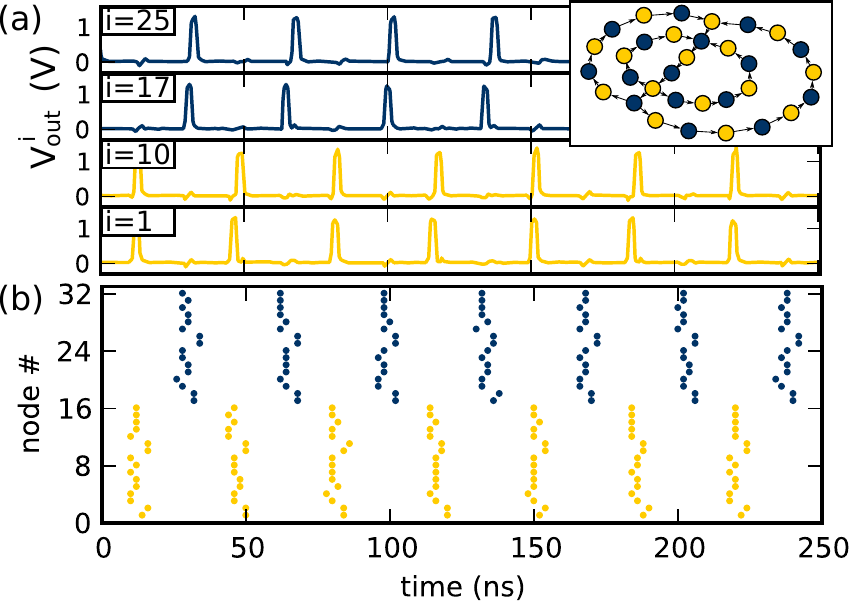}%[width=1\linewidth]
\caption{(Color online) 
Network dynamics of a 2-cluster state with node parameter $n_\mathrm{pulse}=4$ $\left[T_\mathrm{pulse}=(2.2\pm0.1)\un{ns}\right]$, $n_\mathrm{ref}=10$ $\left[T_\mathrm{ref}=(5.6\pm0.2)\un{ns}\right]$ and link delay times $n_\tau=30$ $\left[\tau=(16.7\pm0.6)\un{ns}\right]$ after initial stimulation of one node with one pulse of width $w=(1.6\pm0.1)\un{ns}$. The inset is a replica of the topology of the network, where nodes are colored by cluster.
(a) The output waveform of four nodes in the network. Input-output gates are used for the readout with the oscilloscope ($8\un{GHz}$ analog bandwidth, $40\un{GSa/s}$ sampling rate).
(b) Raster diagram of the network, where each point represents the temporal occurrence of a spike with a $2\un{ns}$ resolution. The network is realized using an Altera Cyclone IV FPGA (EP4CE115F29C7N).
}
\label{fig:2cluster}
\end{figure}

With separated time scales satisfying $\tau>T_\mathrm{ref}>\Delta$, the experimental network displays two zero-lag synchronized clusters as shown in Fig.~\ref{fig:2cluster}a. The waveforms of four nodes, two out of each cluster, show coherent spiking with period $T_\mathrm{cluster}\approx\mathrm{GCD}\cdot\tau=2\tau$. This behavior is also predicted by the GCD theory from the number of elements in each loop, as $\mathrm{GCD}(8,10,12,16)=2$ \cite{KAN11a}. The experimental results are consistent with numerical simulations using a theoretical description of the excitable node introduced in Ref.~\cite{ROS12} (see Supplementary Material).

The spiking dynamics of the entire network is shown in the raster diagram in Fig.~\ref{fig:2cluster}b, where each dot represents a spiking event, subject to a discretization error of $\pm1\un{ns}$. The first (last) 16 elements, as also shown in the inset, are in near zero-lag synchronization and belong to a cluster. The variation ($\pm4\un{ns}$) in spike generation time between nodes is due to differences in the link time delays and measurement error that originates from signal propagation delays on the FPGA. 

To our knowledge, this network is the largest experimentally implemented complex network showing cluster synchronization that operates without computer assistance, which is commonly used to manage the network coupling in experiments \cite{IND11,VAR12,HAG12}. This illustrates that our setup is well-suited to build large networks compared to other experimental approaches \cite{NIX12}.

We can access network dynamics that are not predicted by topological considerations with the GCD by changing the separation of time scales ($\tau>T_\mathrm{ref}>\Delta$) in our system. Time scales are adjusted by changing the refractory time of the excitable nodes. First, we study short refractory times $T_\mathrm{ref}$ on the order of the heterogeneities $\Delta$ of the link time delays and, second, long refractory times $T_\mathrm{ref}$ on the order of the link time delays $\tau$.

\begin{figure}[tb]
\centering
\includegraphics{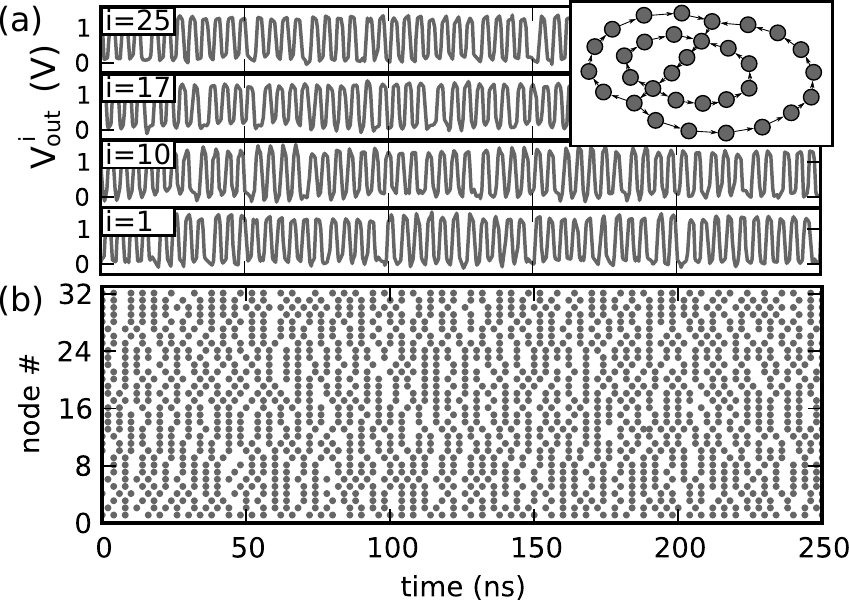}%[width=1\linewidth]
\caption{Same as Fig.~\ref{fig:2cluster}, except with $n_\mathrm{ref}=5$ ($T_\mathrm{ref}=(2.8\pm0.1)\un{ns}$), showing a desynchronized state (synchronization breakdown) of the network dynamics. 
}
\label{fig:breakdown}
\end{figure}

When we decrease the refractory time to a value of $T_\mathrm{ref}=(2.8\pm0.1)\un{ns},$ the network dynamics change. Instead of cluster synchronization with oscillations on the order of $\tau$, the network displays fast, incoherent spiking dynamics with interspike intervals on the order of $T_\mathrm{ref}$, as shown in the waveform and raster diagram of Fig.~\ref{fig:breakdown}. For easy comparison with the previous figure, the time axis is kept the same. The new dynamical state generates excitations constantly, leading to pulsing dynamics with high frequencies close to the maximum frequency allowed by the excitable nodes, given by $1/T_\mathrm{ref}$.

The breakdown is caused by heterogeneity in the loop propagation times at the high-in-degree nodes. With Eq.~\eqref{eq:loop_prop_time}, a maximum time difference $\Delta=\max_{i,j}(\left|\Delta_i-\Delta_j\right|)=(2.8\pm0.1)\un{ns}$ exists in the propagation times $T_i$ of the network loops and leads to a mismatch of the arrival times of pulses. When $\Delta<T_\mathrm{ref}$, the refractory time can compensate for the mismatch, by blocking pulses that arrive a time $\Delta$ after the first pulse during every period of the clusters $T_\mathrm{cluster}$. In this case, the spiking dynamics stays coherent. Otherwise, when $\Delta>T_\mathrm{ref}$, the pulse that arrives a time difference $\Delta$ after the first pulse will trigger additional pulse trains, leading to incoherent high-frequency spiking.

\begin{figure}[tb]
\centering
\includegraphics{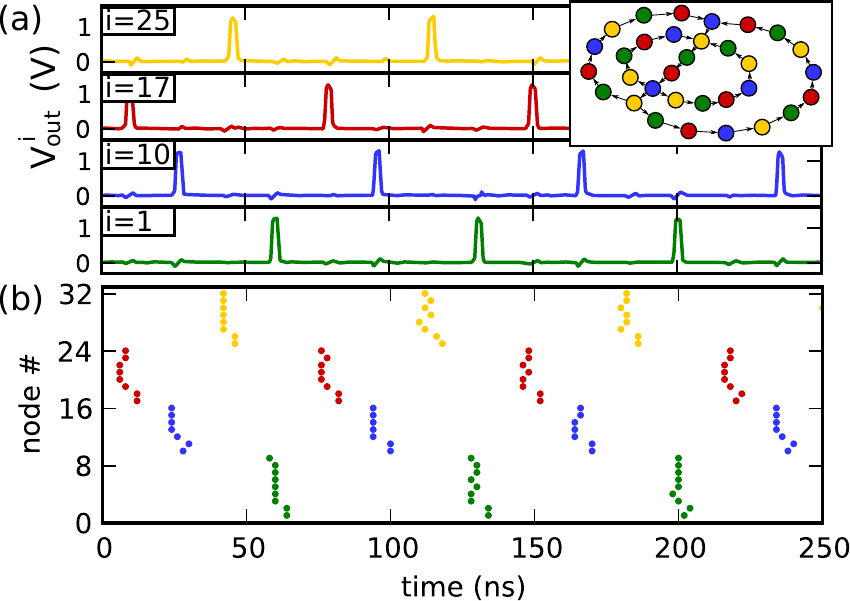}%[width=1\linewidth]
\caption{(Color online) 
Same as Fig.~\ref{fig:2cluster}, except with $n_\mathrm{ref}=70$ ($T_\mathrm{ref}=(39\pm2)\un{ns}$), showing four clusters in zero-lag synchronization (4-cluster state). 
}
\label{fig:4cluster}
\end{figure}

Besides the breakdown for small $T_\mathrm{ref}$, the cluster synchronization patterns are also affected for large $T_\mathrm{ref}$. When the refractory time is increased to $T_\mathrm{ref}=(39\pm2)\un{ns}\approx 2.3\tau$, the network displays four synchronized clusters (4-cluster state) instead of the 2-cluster state that is inferred from the topology and observed in Fig.~\ref{fig:2cluster}, as shown in the waveforms and raster diagram in Fig.~\ref{fig:4cluster}.

To understand this behavior, we consider the maximum output frequency of the excitable node, given by $1/T_\mathrm{ref}$. When the predicted oscillation frequency, given by $1/T_\mathrm{cluster}\approx1/(\mathrm{GCD}\cdot\tau)$, is above the maximum frequency $1/T_\mathrm{ref}$, the network cannot show the predicted cluster state and the dependency on $T_\mathrm{ref}$ comes into effect. Loops that generate pulses with periods $T<T_\mathrm{ref}$, as predicted from the topology, are suppressed at nodes with high in-degree because the pulses fall into the refractory phase of pulses from other loops. Thus, the GCD has to be recalculated by neglecting some loops, using only the loops that lead to a GCD larger than $T_\mathrm{ref}/\tau$. Because this pulse-blocking mechanism is based on short oscillation periods, the loops that are effective for the cluster dynamics and used for the calculation are those that lead to the smallest value of the GCD that is greater than $T_\mathrm{ref}/\tau$, \emph{i.e.}, that lead to the shortest period greater than $T_\mathrm{ref}$. Therefore, with the size of loops $L_i$, the number of clusters is given by
\begin{equation}\label{eq:GCD_crit}
\min_{\{L_i\}\in \text{Network}}\left[ \mathrm{GCD}(\{L_i\})\right] : \mathrm{GCD}(\{L_i\})>T_\mathrm{ref}/\tau.
\end{equation}

This extended criterion describes successfully the stable synchronization patterns observed in Fig.~\ref{fig:4cluster}. Because $\mathrm{GCD}(8,10,12,16)=2<T_\mathrm{ref}/\tau\approx2.3$, a loop needs to be exempt from the calculation, leading to the next larger GCD value of $\mathrm{GCD}(8,12,16)=4>T_\mathrm{ref}/\tau\approx2.3$. Therefore, the modified GCD theory predicts the experimentally observed 4-cluster state.

The constraint given by $T_\mathrm{ref}$ depends on the network topology. For example, a network with predicted zero-lag synchronization (1-cluster state) transitions to a different cluster state already when $T_\mathrm{ref}/\tau>1$. When $T_\mathrm{ref}$ is further increased, more and more loops lose their effect until the refractory time is longer than the propagation delay through the largest loop; then, spiking dynamics is no longer self-sustained and the network relaxes to the quiescent state. Surprisingly, in our topology, it is not the shortest loop with eight nodes but the loop with ten nodes that becomes ineffective first.

\paragraph{Control of synchronization patterns}
A global adjustment of the refractory time $T_\mathrm{ref}$ of all nodes influences the network dynamics significantly. However, we can even achieve local control of the global network dynamics by adjusting the properties of only a small number of nodes.

The influence of the refractory time on the network dynamics is most prominent at the nodes with high in-degree. This motivates us to only adjust the refractory times of the nodes with in-degree greater than one, which represents a simple degree-correlation \cite{BRE08a}.

First, we investigate the network dynamics for short refractory times. We set the refractory times of nodes to $T_\mathrm{ref}=(2.8\pm0.1)\un{ns}$, a value for which the breakdown of cluster synchronization is observed. When we now increase the refractory times of the two nodes with an in-degree greater than one (node 13 and 30 in Fig.~\ref{fig:setup}b)  to $T_\mathrm{ref}=(5.6\pm0.2)\un{ns}$, the stable synchronization patterns reappear as a solution. An initial pulse sent to this network in the quiescent state leads to a 2-cluster synchronization pattern similar to that observed in Fig.~\ref{fig:2cluster}.

Second, we investigate network dynamics for large refractory times. We set the refractory times of all nodes to $T_\mathrm{ref}=(5.6\pm0.2)\un{ns}$, a value for which a 2-cluster synchronization state is observed. When we now increase the time scales of the two nodes with an in-degree greater than one to $T_\mathrm{ref}=(39\pm2)\un{ns}$, the stable synchronization pattern changes to a 4-cluster state, which we have observed in Fig.~\ref{fig:4cluster}, when increasing the time scales of all nodes.

Both cases allow for the control of the synchronization patterns locally by a small fraction of the network nodes by adjusting the refractory time of only 2 out of 32 nodes.

\paragraph{Conclusion}
Cluster synchronization patterns change when the refractory time of the nodes is larger than the link time delays or smaller than the heterogeneity of the link time delays. For large refractory times, cluster synchronization patterns are modified, and, for short refractory times, cluster synchronization breaks down to an incoherent state. In both cases, we identify the mechanism leading to the transition and, in the first case, we put forth a modified GCD criterion that includes the constraints imposed by the refractory time. The synchronization patterns can be controlled by the refractory time of a small fraction of nodes, identified by their in-degree. 

In addition to the potential application to neuroscience stated in the introduction, our findings have two fundamental implications. First, the dynamics of neural networks does not solely depend on the global topology as suggested by Kanter~et~al.~\cite{KAN11a}. But, depending on the time scale of the nodes, some links are dynamically pruned, leading to a new effective topology with altered synchronization patterns, as described by Eq.~\eqref{eq:GCD_crit}. Second, the driver nodes relevant for control can be identified easily by their large in-degree and allow one to control the global network dynamics locally. This is similar to a recent study on the controllability of networks that predicts that the number of driver nodes is given by the network's degree distribution~\cite{LIU11}.

%acknowledgement, does not need to be counted for the words
D.P.R., D.R., and D.J.G. gratefully acknowledge the financial support of the U.S. Army Research Office Grant W911NF-12-1-0099. E.S. and D.P.R. acknowledge support by the DFG in the framework of SFB~910. We thank Forrest Friesen, Seth Cohen, and Kristine Callan for fruitful discussions.

  %\bibliography{references}
  %\bibliography{ref}
%
\end{document}